\documentclass[a4paper, 12pt]{article}
\usepackage[version=3]{mhchem} 
\usepackage[T1]{fontenc}     
\usepackage{amssymb}
\usepackage{amsmath}
\usepackage{amsfonts}
\usepackage{amsthm}
\usepackage{fancyhdr}
\usepackage{amsbsy}
\usepackage{gensymb}
\usepackage{epstopdf}
\usepackage{graphicx}
\usepackage{authblk}
\usepackage{cite}
\usepackage{color}
\usepackage{bbding}
\usepackage[section]{placeins}
\usepackage{float}

\begin{document}
\title{\textbf{Crystal Structure Elucidation of the Novel Molecular-Inorganic Polymer $\mathrm{K_{2}SeO_{4}\cdot H_{2}SeO_{3}}$.}}   

\author[1,2]{Oscar S. Hern{\'a}ndez-Daguer \Envelope} 
\author[3]{Charles L. Barnes}
\author[4]{Samuel P. Hern{\'a}ndez-Rivera}
\author[5]{Jorge L. R{\'i}os-Steiner \Envelope} 
\affil[1]{Department of Physics, University of Massachusetts, Amherst, Massachusetts 01003, United States, e-mail: ohernandezda@umass.edu, oscarsaidh@gmail.com}
\affil[2]{Department of Physics, University of Puerto Rico, Mayag{\"u}ez, Puerto Rico 00681-9000, United States, e-mail: oscar.hernandez3@upr.edu, oscarsaidh@gmail.com, phone: +1 787 228 1380.}
\affil[3]{Department of Chemistry, University of Missouri, Columbia, Missouri 65211, United States.}
\affil[4]{ALERT-II DHS Center of Excellence for Explosives Research, Department of Chemistry, University of Puerto Rico, Mayag{\"u}ez, Puerto Rico, 00681-9000, United States.}
\affil[5]{Laboratory of Crystallography and Synthesis of New Materials, Department of Chemistry, University of Puerto Rico, Mayag{\"u}ez, Puerto Rico, 00681-9000 United States, e-mail: jorge.rios2@upr.edu, phone: +1 787 832 4040 ext 2538, Fax: +1 787 265 3849.}
\setcounter{Maxaffil}{0}
\renewcommand\Affilfont{\itshape\footnotesize}
\maketitle

\begin{scriptsize}
\begin{abstract}
The orthorhombic crystal structure of the novel molecular-inorganic polymer $\ce{K2SeO4\cdot H2SeO3}$ was elucidated using single-crystal X-ray diffraction with MoK$\alpha$ radiation ($\lambda$= 0.71073 {\AA}), performed at 100 and 298.15 K. The reported data is at 100 K since there were no structural differences as compared to the room temperature. The technique revealed $\ce{K2SeO4\cdot H2SeO3}$ crystals to have space group $Pbcm$ with unit cell dimensions $a =$ 8.8672(17) {\AA}, $b =$  7.3355(14) {\AA}, $c =$ 11.999(2) {\AA} and $Z =$ 4. The unit cell volume obtained was $V =$ 780.5(3) {\AA}$^3$ with a calculated density $D_{c} = 2.980 Mg/m^3$. In $\ce{K2SeO4\cdot H2SeO3}$, the selenate anions and selenous acid molecules form an infinite polymeric chain along the \textit{c} axis, through strong hydrogen bonds. That is to say; there are two distinctive species forming a polymeric sequence extended along the \textit{c} axis with an alternating molecule-anion-molecule $\mathrm{(SeO_{4}^{2-}-H_{2}SeO_{3})_{n}}$, defined as a Molecular Inorganic Polymer (MIP).  The $\mathrm{K^{+}}$ cations influence the orientation of the $\mathrm{SeO_{4}^{2-}}$ anions, and these consequently affect the arrangement of the $\mathrm{H_{2}SeO_{3}}$ molecules within the structure.  The crystal packing forces are governed by ionic and dipole-dipole interaction.  Full-matrix least-square refinement method on $F^2$ provide final Reliability indices of $R_{1}$ = 0.0150, and $wR_{2}$ = 0.0377, and a Goodness-of-fit = 1.140; at 0.75 {\AA} resolution, where $R_{int}$ = 0.0230, and $F(000)$ = 656 for 1014 independent reflections ($I>2\sigma(I)$), from a total of 8,576 reflections collected.
	Based on the crystallographic similarities of this novel structure with other previously reported compounds, potential non-linear optical, fast-proton conduction and acid transporting host material properties can be expected to develop solid-state technological applications with this crystal.
\end{abstract}

\end{scriptsize}
\section{Introduction}
The coupling of these two intricate and concerning issues, global warming \cite{Tollefson2018, Christidis2011, Monastersky2009, Arrhenius1896}(that is requiring drastic actions \cite{Tollefson2018})and the energy crisis\cite{Newton2012, T.M.Rybczynski1976}, have motivated the direction of most research efforts towards the development of alternative energy sources. The efforts to develop these alternatives \cite{Radenahmad2016, AGRAWAL1999} and the development of materials with high ionic conduction, also known as Superionic Conductors, have encouraged scientists to study materials with high ionic conductive properties.  Since the 1970’s, study in this field has developed rapidly \cite{Hoshino1991}, mainly due to the potential applications in fuel cell technology.  Fuel cells are electrochemical devices that use electrolytes to produce electricity.  It has become an energy alternative (along with solar energy) since it generates electricity with high efficiency by combining hydrogen and oxygen electrochemically without combustion, producing water and heat as by-products \cite{Tollefson2010}.
  
Polymer electrolyte membranes (PEM) are currently considered the most adequate for the construction of portable fuel cells, making it ideal for automotive applications, like cars \cite{Tran2015, Liu2016, Service2004}, trains \cite{Jafri2017, Hoffrichter2014, InstitutionofEngineeringandTechnology.2016a} and aircraft \cite{Guida2017, Sarlioglu2015, Kivits2010}.  These PEM Fuel cells (PEMFCs) have many advantages, such as rapid starting time, high energy efficiency and compactness, among others.  Despite this, membranes require the continuous addition of water to transfer the protons through the electrolyte solution and also require platinum to catalyze the oxidation of $\mathrm{H_{2}}$ and the reduction of $\mathrm{O_{2}}$ \cite{Tran2015} his restricts the membranes to have to operate below the boiling point of water, the temperature range at which precious-metal catalysts work slowly and can become inactivated by the binding of carbon monoxide\cite{Service2004}. Different types of fuel cells have been proposed to overcome these issues, such as  Alkaline Fuel Cells (AFCs) \cite{Shui2015}, Protonic Ceramic Fuel Cells (PCFCs) \cite{Gorte2015}, and one of the most studied, solid Acid Fuel Cells (SAFCs) \cite{Service2004}.

The evidence of a superprotonic phase in solid acid salts, such as in the family of $\mathrm{MHAO_{4}}$ , $\mathrm{M_{3}H(AO_{4})_{2}}$ , $\mathrm{M_{4}H_{2}(AO_{4})_{3}}$, $\mathrm{M_{5}H_{7}(RO_{4})_{4}}$, $\mathrm{M_{5}H_{3}(AO_{4})_{4}\cdot xH_{2}O}$, $\mathrm{M_{9}H_{7}(AO_{4})_{8}\cdot H_{2}O}$ (M = K$^{+}$, Rb$^{+}$, $\mathrm{(NH_{4})^{+}}$, Cs$^{+}$; A = S, Se), and $\mathrm{MH_{2}RO_{4}}$ (R= P, As) \cite{Baranov2003,Rhimi2018a,Gaydamaka2018} and other compounds created based on these proton conductor families \cite{Belghith2015, Yoshii2015, Lee2015, Litaiem2015, Weil2016, Chisholm1999, Ortiz2007, Haile1997, Boubia1985, Ghorbel2015, Makarova2016, Nouiri2016, Qing2016, Ikeda2014, Ponomareva2011, Selezneva2018} , opens the possibility for a wider range of temperatures in which the fuel cells operate (outside the temperature range in which the water is liquid, assuming that materials with a superprotonic phase would be developed below 0 $\degree$C).  In addition, these solid acid electrolytes might reduce manufacturing costs by eliminating the use of platinum as a catalyst, without disregarding the possibility for yields and lifetime increases for these devices \cite{AGRAWAL1999}.
Compounds of the $\mathrm{M_{3}X(AO_{4})_{2}}$ family \cite{Lindner2017a, Hatori2006}, such as $\mathrm{K_{3}H(SeO_{4})_{2}}$, $\mathrm{Rb_{3}H(SeO_{4})_{2}}$, $\mathrm{(NH_{4})_{3}H(SeO_{4})_{2}}$, etc., possess high proton conductivity and have been broadly reported to have a superionic phase transition above 112 $\degree$C \cite{Wolak2013, Hatori2006, Pawowski2007, Shikanai2011, Shikanai2010}, encouraging the design of fuel cells that operate at temperatures above the normal boiling point of water.
However, some inconsistencies have been found in the thermal and electrical behavior of members of the $\mathrm{M_{3}X(AO_{4})_{2}}$ family \cite{Pawlowski2003, Pawowski2007, MATSUO2007, Chen2002, Al-kassab1993} and other families mentioned above.  Due to these inconsistencies, different physical and chemical points of view have attempted to explain the process that produces the fast ion conduction in $\mathrm{M_{3}X(AO_{4})_{2}}$ acid salts \cite{Baranov1988, Simonov2009, Hatori2006, Chen2002, COWAN2008, Chisholm2001, Chen2003, Poomska2003, Ichikawa1993, DevendarReddy1982, Ito1998, Shikanai2010, Wolak2013, Pavlenko1999, Shikanai2009, Yoshida2012, Gordon1995, Merunka2009, Pavlenko2011a, Suzuki2006, Bednarski2008, Sohn2013a, Makarova2010, Mkarova2011a}, as is the case of $\mathrm{K_{3}H(SeO_{4})_{2}}$ (abbreviated as TKHSe).  The superionic phase transition in TKHSe has been observed around $T_{p}$ (114 $\degree$C) by performing thermal and electrical studies under different experimental tests and conditions.  The results published by different authors reflect discrepancies in the superionic phase transition temperature $T_{p}$ values.  Values have been reported in the range between 108.1 $\degree$C and 122 $\degree$C \cite{Al-Kassab1993, Matsumoto2001a, Yokota1982, Chen2002}.  A recent report \cite{Hernandez-Daguer2015a} claims  that TKHSe  undergoes a phase transition at around 112 $\degree$C and simultaneously decomposes in the temperature range of 110 to 150 $\degree$C, starting at the surface of the TKHSe grains. The assumption is that the decomposition is analogous to that reported by Lee \cite{Lee1996}, for the KDP-type crystals, which suggest that the observed decrease in conductivity on successive thermal runs is a consequence of thermal decomposition (thermally activated).  However, the jump in conductivity is only a consequence of the order-disorder transition in the TKHSe phase that remains inside the grains.  This implies that TKHSe cannot fulfill the electrical and thermal requirements to be applied in fuel cells; due to thermal instability in its superprotonic phase.  
O.S. Hernández-Daguer et al. \cite{Hernandez-Daguer2015a} , presumes that through variations in the atomic proportions in the $\mathrm{M_{3}H(AO_{4})_{2}}$ family, that is, by increasing the amount of selenate tetrahedra, one could possibly generate new chemical structures and materials that preserve the electrical properties and lack the thermal instability present in $\mathrm{K_{3}H(SeO_{4})_{2}}$ (analogous to thermal stability profiles present in ceramic materials \cite{Hernandez2009a}).
Based on this hypothesis and looking for the superionic phase transition temperatures of the $\mathrm{M_{4}H_{2}(AO_{4})_{3}}$ family \cite{Al-kassab1993Rb4, Hilczer1999, Pawowski2001, Poomska2000, Fukami1999a, Hilczer1999, Augustyniak1992}, we were able to synthesize a novel compound, $\mathrm{K_{2}SeO_{4}\cdot H_{2}SeO_{3}}$. The chemical structure of this compound was determined by single crystal X-ray diffraction.

\section{Experimental} 
In order to synthesize some selenic acid salts, one of the products obtained was a novel compound, identified using single crystal X-ray diffraction and selected for our studies. Single crystals of the new compound were selected under a polarizing microscope (for size and optical quality selection purposes – in order to avoid multiple aggregates, twinned crystals and to improve diffraction quality) and recovered with paratone oil, to avoid the hydration of the crystals, and mounted on a Metegen micropin.  The chosen crystals were 100 to 500 $\mu$m in size.  Diffraction data for the novel compound was collected at 293K and at 100K with a Bruker diffractometer equipped with a graphite monochromator using MoK$\alpha$ radiation ($\lambda$= 0.71073 {\AA}).  This equipment is also coupled to an APEX II CCD area detector system and an Oxford Cryostream Low Temperature device.  Data was collected using the $\omega$ and $\phi$ scan techniques.  The structure and data collected at low temperature is the only one that will be reported since no significant structural differences were observed when compared to the room temperature structure.  The temperature factors data is also better at low temperature, since the atoms are more localized due to less vibration in the crystal, which also improve the intensities and shape of the spots.  Preliminary orientation matrixes (orientation matrix that specifies the orientation of the crystal and detector with respect to the X-ray beam), and unit cell parameters were obtained from $\phi$ and $\omega$ scans between 0$\degree$ $\leqslant$ $\phi$ $\leqslant$ 90$\degree$, 20$\degree$ $\leqslant$ $\omega$ $\leqslant$ 70$\degree$ and then refined using the whole data set.  Frames were integrated and corrected for Lorentz and polarization effects.  Equivalent reflections were merged, and absorption corrections were made using an empirical absorption methods \cite{Gorbitz1999a}, with $\mathrm{T_{min}=}$ 0.11 and $\mathrm{T_{max}}=$ 0.33 (Max. and Min. transmission). Space group, lattice parameters and other relevant information are listed in Table~\ref{tbl:cdsrphK}.
\begin{table}
\fontsize{10}{12}\selectfont
\centering
  \caption{Crystal data and structure refinement parameters for $\ce{K2SeO4\cdot H2SeO3}$}
  \label{tbl:cdsrphK}
  \tabcolsep=0.11cm
  \begin{tabular}{lll}
    \hline
     &	& \\
    \hline
    Identification code   & $\ce{K2SeO4\cdot H2SeO3}$	& \\
    Empirical formula    & $\ce{H2 K2 O7 Se2}$		&   \\
    Formula weight  & 350.14	&  \\
    Temperature  & 100(2) K	&   \\
    Wavelength & 0.71073 {\AA}	& \\
    Crystal system  & Orthorhombic	&\\
    Space group  & $Pbcm (N\degree 57)$	&\\
    Unit cell dimensions 	& $a =$ 8.8672(17) {\AA} & $\alpha = 90\degree$ \\
     						& $b =$  7.3355(14) {\AA}	& $\beta = 90\degree$ \\
     						& $c =$ 11.999(2) {\AA}	& $\gamma = 90\degree$ \\
    Volume	 & 780.5(3) {\AA}$^3$	& \\
    Z & 4	& \\
    Density (calculated)	 & 2.980 Mg/m$^{3}$	& \\
    Absorption coefficient & 10.532 mm$^{-1}$	& \\
    $F(000)$  & 656	& \\
    Crystal size	 & 0.500 x 0.250 x 0.200 mm$^{3}$	&  \\
    Theta range for data collection & 2.297\degree $\leq \theta \leq$  28.676\degree.	& \\
    Index ranges	& $-11 \leq h \leq 11$,& \\
    				& $-9 \leq k \leq 9$, & \\
    				& $-15 \leq l \leq 16$	& \\
    Reflections collected & 8576	& \\
    Independent reflections & 1014 [$R_{(int)}$ = 0.0230] & \\
    Completeness to theta = 25.242\degree	& 100.0 \% & \\
    Absorption correction	& Semi-empirical from equivalents	& \\
    Max. and min. transmission &	0.23 and 0.11	& \\
    Refinement method &	Full-matrix least-squares on $F^{2}$ & \\	
    Data / restraints / parameters  &	1014 / 0 / 61 & \\
    Goodness-of-fit on $F^{2}$ & 1.140 & \\
    Final $R$ indices [$I>2\sigma(I)$] & $R_{1} = 0.0150$, $wR_{2} = 0.0377$ & \\
    $R$ indices (all data) & $R_{1} = 0.0159$, $wR_{2} = 0.0380$ &\\
    Extinction coefficient	& 0.0093(4) & \\
    Largest diff. peak and hole & 0.575 and -0.458 e.{\AA}$^{-3}$ & \\
    \hline
  \end{tabular}
\end{table} 

The structure was solved by direct methods and refined by fullmatrix least-squares on $|F|^{2}$ (SHELX 97 program \cite{Sheldrick2008}), with the aid of the program X-SEED \cite{Barbour2001}.  Anisotropic thermal factors were assigned to all the non-hydrogen atoms.  All the hydrogen atoms were located in the $F_{o}-F_{c}$ map and refined accordingly.  The fullmatrix least-squares methods minimizes the function $w(F_{o} - F_{c})^{2}$ , where $w$ represents the weight to be assigned an observation, associated to this discrepancies between the observed and calculated values of $F^{,}s$ (or $|F|^{2}$).  The final refined atomic parameters are given in Table~\ref{tbl:achk}.
\begin{table}
\centering
  \caption{Atomic coordinates (x$10^{4}$) and equivalent isotropic displacement parameters ({\AA}$^{2}$ x$10^{3}$) for $\ce{K2SeO4\cdot H2SeO3}$.  $U_{(eq)}$ is defined as one third of the trace of the orthogonalized $U_{ij}$ tensor.}
  \label{tbl:achk}
  \tabcolsep=0.11cm 
  \begin{tabular}{lllll}
    \hline
      & x & y & z & $U_{(eq)}$  \\
    \hline
    Se(1) 	& 8184(1)	& 7500 		& 5000 		& 8(1) 	\\
	Se(2) 	& 3742(1)	& 5921(1) 	& 2500 		& 11(1) \\
	K(1) 	& 7938(1)	& 2500 		& 5000 		& 11(1) \\
	K(2) 	& 10728(1)	& 4584(1) 	& 7500 		& 10(1) \\
	O(1) 	& 7132(2)	& 5928(2) 	& 4401(1) 	& 16(1) \\
	O(2) 	& 9236(2) 	& 6548(2) 	& 5952(1) 	& 14(1) \\
	O(3) 	& 4574(2) 	& 7086(2) 	& 3610(1) 	& 16(1) \\
	O(4) 	& 2121(2) 	& 6974(3) 	& 2500		& 12(1) \\
    \hline
  \end{tabular}
\end{table}

 \section{Results and Discussion} 
\begin{figure}[!htb]
 \begin{center}
  \includegraphics[scale=0.65]{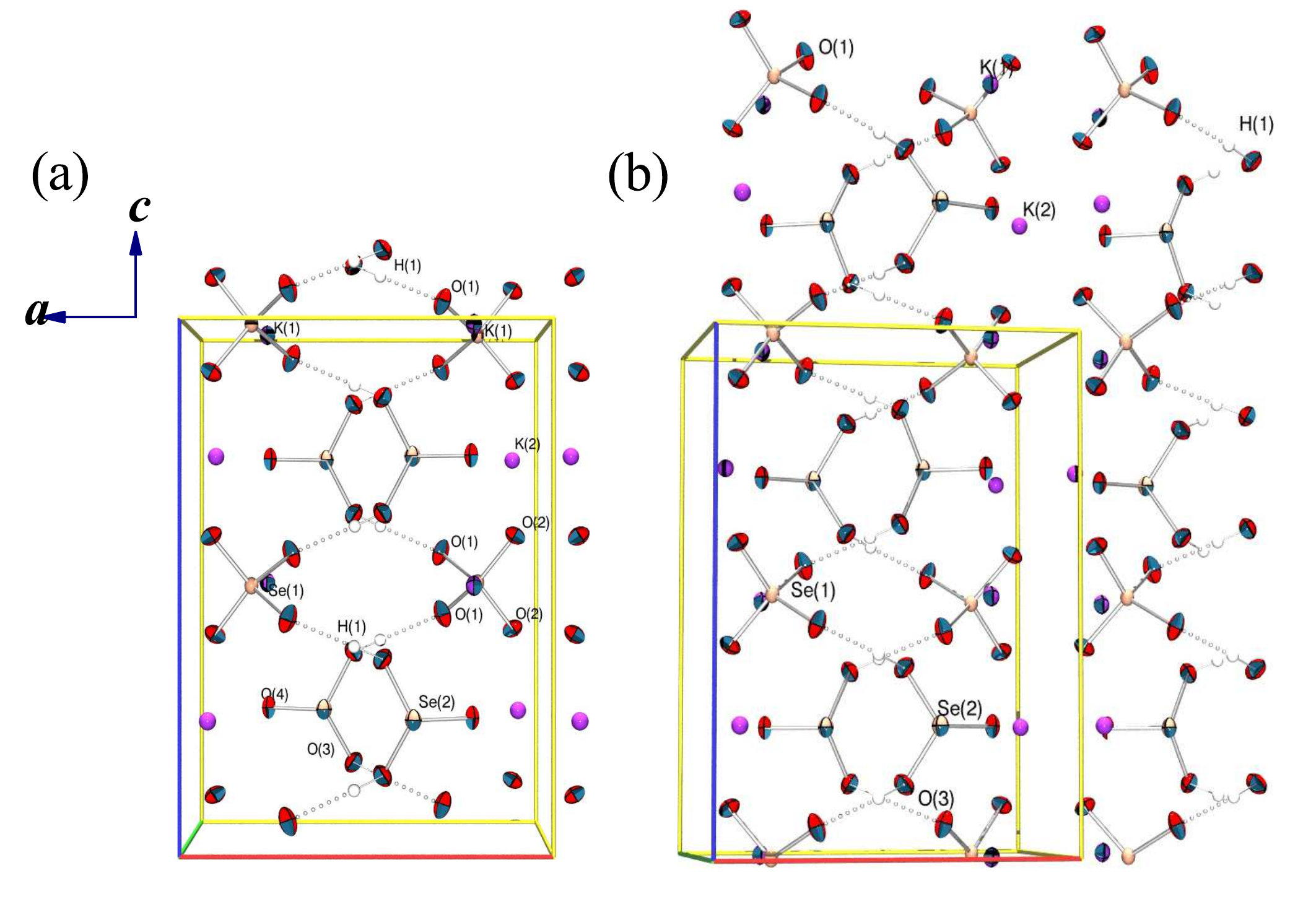}
   \caption{The crystal structure of $\ce{K2SeO4\cdot H2SeO3}$ at 100K.  (a) Projection of the crystal structure onto the $ac$ plane or (010).  (b) Clinographic projection of the structure.  The dot lines in the figure represent the hydrogen bonds.  The figure was generated with the crystal structure visualization program Ortep III \cite{Farrugia2012} and with the aid of the program POV-Ray \cite{ThePOV-Ray2000}.}
  \label{fig:101Clino}
 \end{center}
\end{figure}

Figure~\ref{fig:101Clino} displays a visualization of two different views of the suitable trial structures model found for the crystal structure at $\ce{K2SeO4\cdot H2SeO3}$, which is a projection on $ac$ plane.  The dotted lines in the figure represent hydrogen bonds and solid lines indicate covalent bonding.  The ellipsoids (so called thermal ellipsoids) in Figure~\ref{fig:101Clino} show the amount by which atoms are displaced in a given direction (showed by the shape of the ellipsoid, a cigar shape indicating much motion or displacement). And also indicates the direction of maximum motion \cite{PickworthGlusker1998}.  The $\ce{K2SeO4\cdot H2SeO3}$ crystal is built of potassium cation environments, selenate anions and molecules of selenous acid, similar to $\ce{Na2SeO4\cdot H2SeO3\cdot H2O}$ \cite{Baran1991}.  The structure consists of slightly distorted tetrahedron $\mathrm{SeO_{4}^{2-}}$ anions, electrically equilibrated by $\mathrm{K^{+}}$ cations, and joined by hydrogen bonds with an almost regular pyramidal $\ce{SeO3}$ anion to form a double sequence of anion-molecule-anion extending parallel to the $c$ axis, see Figure~\ref{fig:101Clino}.

\begin{figure}[htb]
 \begin{center}
  \includegraphics[scale=0.45]{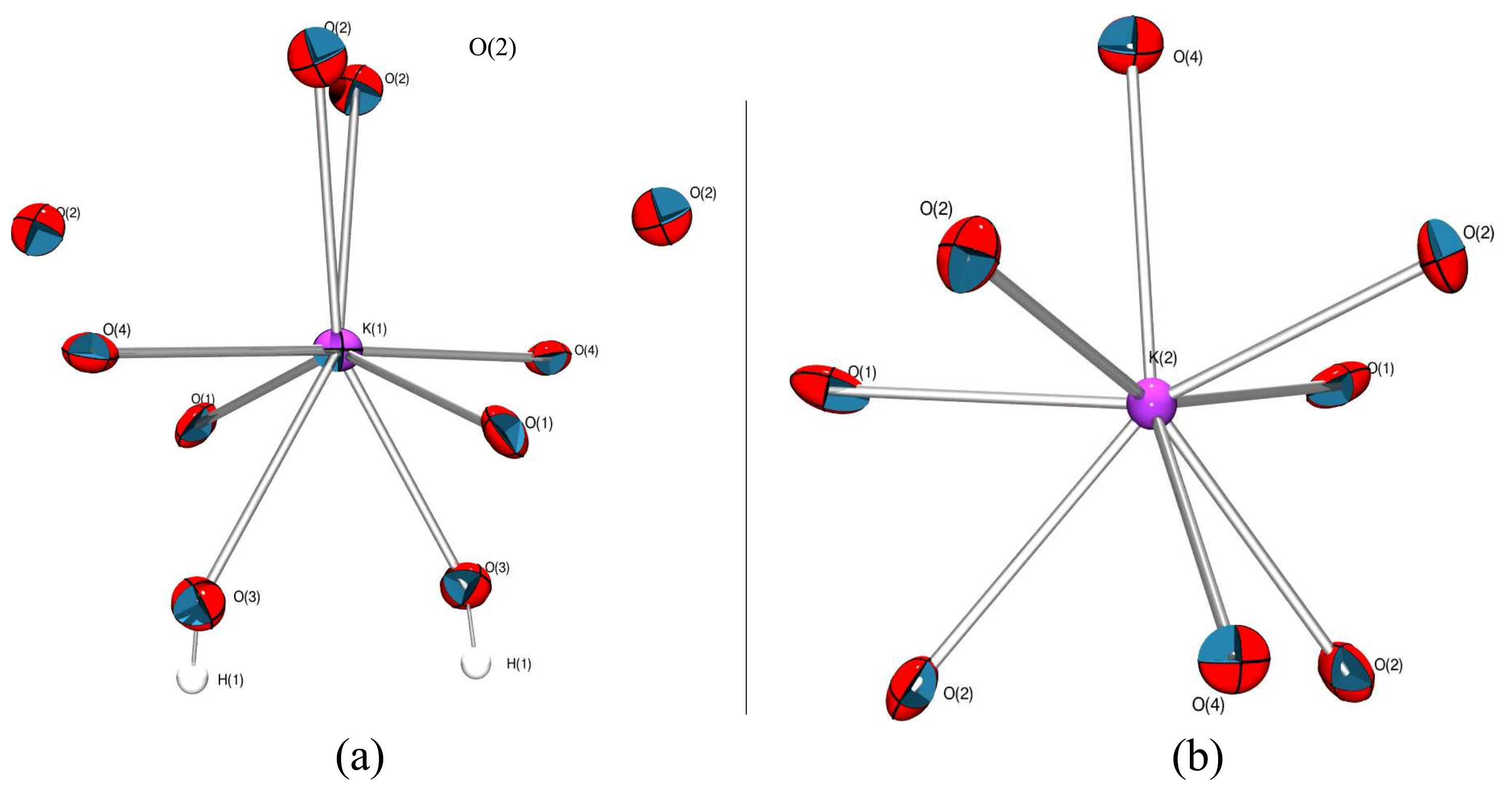}
   \caption{The potassium atoms environment.  (a) Oxygen atoms surrounding K(1) and (b) oxygen atoms around K(2). The figure was generated with the crystal structure visualization program Ortep III \cite{Farrugia2012} and with the aid of the program POV-Ray \cite{ThePOV-Ray2000}.}
  \label{fig:K1K2environment}
 \end{center}
\end{figure}

The crystals of $\ce{K2SeO4\cdot H2SeO3}$ have two types of K cations (See Figure~\ref{fig:K1K2environment} and Table~\ref{tbl:K-O distances}): K(1) occupies special positions and is surrounded by ten oxygen atoms, see Figure~\ref{fig:K1K2environment} a, of which eight are coordinated to it.  Two of them are at a longer distance, 3.3835 \AA, but they seem to be part of the environment of K(1).  K(2) is at a general position and is coordinated to eight oxygen atoms (Figure~\ref{fig:K1K2environment} b).

\begin{table}[H]
\centering
  \caption{Potassium-oxygen interatomic bonds distances, in $\ce{K2SeO4\cdot H2SeO3}$ crystal structure.}
  \label{tbl:K-O distances}
  \tabcolsep=0.11cm
  \begin{tabular}{llll}
    \hline
	Bond									& Distance (\AA)	& Bond							 & Distance (\AA) \\
	\hline
K(1)-O(1)\textsuperscript{\emph{\#8}} 	& 2.7109(15)	& K(2)-O(2) 							 & 2.6977(14) \\
K(1)-O(1) 								& 2.7109(15)	& K(2)-O(2)\textsuperscript{\emph{\#11}} & 2.6977(14) \\
K(1)-O(3)\textsuperscript{\emph{\#9}} 	& 2.7987(16)	& K(2)-O(4)\textsuperscript{\emph{\#7}}	 & 2.773(2)	 \\
K(1)-O(3)\textsuperscript{\emph{\#7}} 	& 2.7987(16)	& K(2)-O(4)\textsuperscript{\emph{\#12}} & 2.811(2) \\
K(1)-O(2)\textsuperscript{\emph{\#10}}	& 2.8409(16)	& K(2)-O(2)\textsuperscript{\emph{\#10}} & 2.9002(15) \\
K(1)-O(2)\textsuperscript{\emph{\#2}}	& 2.8409(16)	& K(2)-O(2)\textsuperscript{\emph{\#13}} & 2.9002(15) \\
K(1)-O(4)\textsuperscript{\emph{\#9}}	& 3.0248(6)		& K(2)-O(1)\textsuperscript{\emph{\#14}} & 2.9907(16) \\
K(1)-O(4)\textsuperscript{\emph{\#7}}	& 3.0248(6)		& K(2)-O(1)\textsuperscript{\emph{\#2}}	 & 2.9907(16) \\
K(1)-O(2)\textsuperscript{\emph{\#8}}	& 3.3834(15)	& \textbf{Average} 						 & 2.8451	 \\
K(1)-O(2) 								& 3.3835(15)	&										 &			 \\			
\textbf{Average} 						& 2.9517		&										 &			 \\
\hline
\end{tabular}
  
\textsuperscript{\emph{\#2}}: -x+2,-y+1,-z+1  \textsuperscript{\emph{\#7}}: -x+1,-y+1,-z+1  \textsuperscript{\emph{\#8}}: x,-y+1/2,-z+1  \textsuperscript{\emph{\#9}}: -x+1,y-1/2,z \textsuperscript{\emph{\#10}}: -x+2,y-1/2,z  \textsuperscript{\emph{\#11}}: x,y,-z+3/2  \textsuperscript{\emph{\#12}}: x+1,-y+3/2,-z+1  \textsuperscript{\emph{\#13}}: -x+2,y-1/2,-z+3/2  \textsuperscript{\emph{\#14}}: -x+2,-y+1,z+1/2
\end{table}

\begin{figure}[!htb]
 \begin{center}
  \includegraphics[scale=0.40]{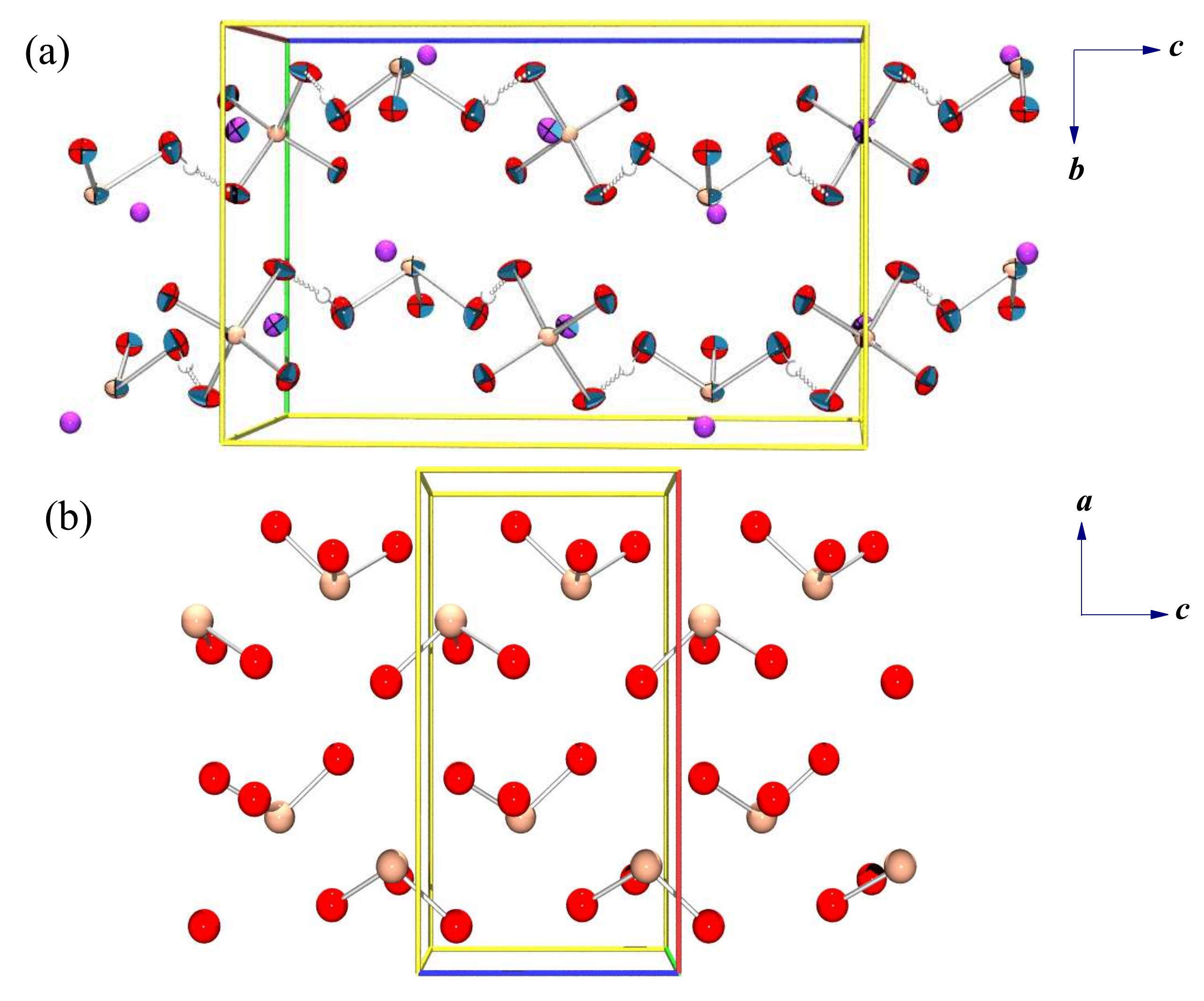}
   \caption{Representation of: (a) polymeric anion-molecule-anion $ \mathrm{(SeO_{4}-H_{2}SeO_{3})_{n}}$ sequence, projected onto the $bc$ plane for $\ce{K2SeO4\cdot H2SeO3}$ crystal structure. (b) sequence $\mathrm{(H_{2}SeO_{3}-H_{2}SeO_{3})}$ onto the $ac$ plane \cite{Wells1949}.  The figure was generated with the crystal structure visualization program Ortep III \cite{Farrugia2012} and with the aid of the program POV-Ray \cite{ThePOV-Ray2000}.}
  \label{fig:comparison}
 \end{center}
\end{figure}

	However, both types of potassium atoms have \textit{eight} oxygen atoms as their nearest neighbors, with an average potassium-oxygen distance of 2.8438 and 2.8451{\AA} for K(1) and K(2) respectively.  Including all oxygen atoms surrounding the potassium atoms the average potassium-oxygen distance are 2.9517 {\AA} for K(1) and 2.8451 {\AA} for K(2),  unlike the average distances and the number of surroundings oxygen atoms reported for $\ce{K2SeO4}$ \cite{Kalman1970}, with average values of 3.14 {\AA} for K(1) and 2.93 {\AA} for K(2).
	The Se(1) is  at a special position and Se(2) occupies a general position (Table~\ref{tbl:achk} and  Figure~\ref{fig:101Clino}).  The crystal packing shows selenate anions and selenous acid molecules form an infinite polimeric chain of alternating species along the $c$ axis, mediated by strong hydrogen bonds, salt bridges and ion-dipole interactions \cite{Jeffrey1999}.  In other words, $ \mathrm{(SeO_{4}^{2-}-H_{2}SeO_{3})_{n}}$, defined as a \textit{Molecular Inorganic Polymer} (MIP), see Figure~\ref{fig:comparison}a, seems to resemble the selenous acid crystal structure \cite{Wells1949} as shown in Figure~\ref{fig:comparison}.  The $\ce{H2SeO3}$ and $\mathrm{SeO_{4}^{2-}}$ anions, in $\ce{K2SeO4\cdot H2SeO3}$ crystal structure, are oriented towards the $\mathrm{K^{+} }$ ions.  The $\mathrm{K^{+} }$ ions, with their electrical charge, influence the orientation of the $\mathrm{SeO_{4}^{2-}}$ anions and these consequently affect the arrangement of the $\ce{H2SeO3}$ molecules within the structure.

\begin{figure}[!htb]
 \begin{center}
  \includegraphics[scale=0.30]{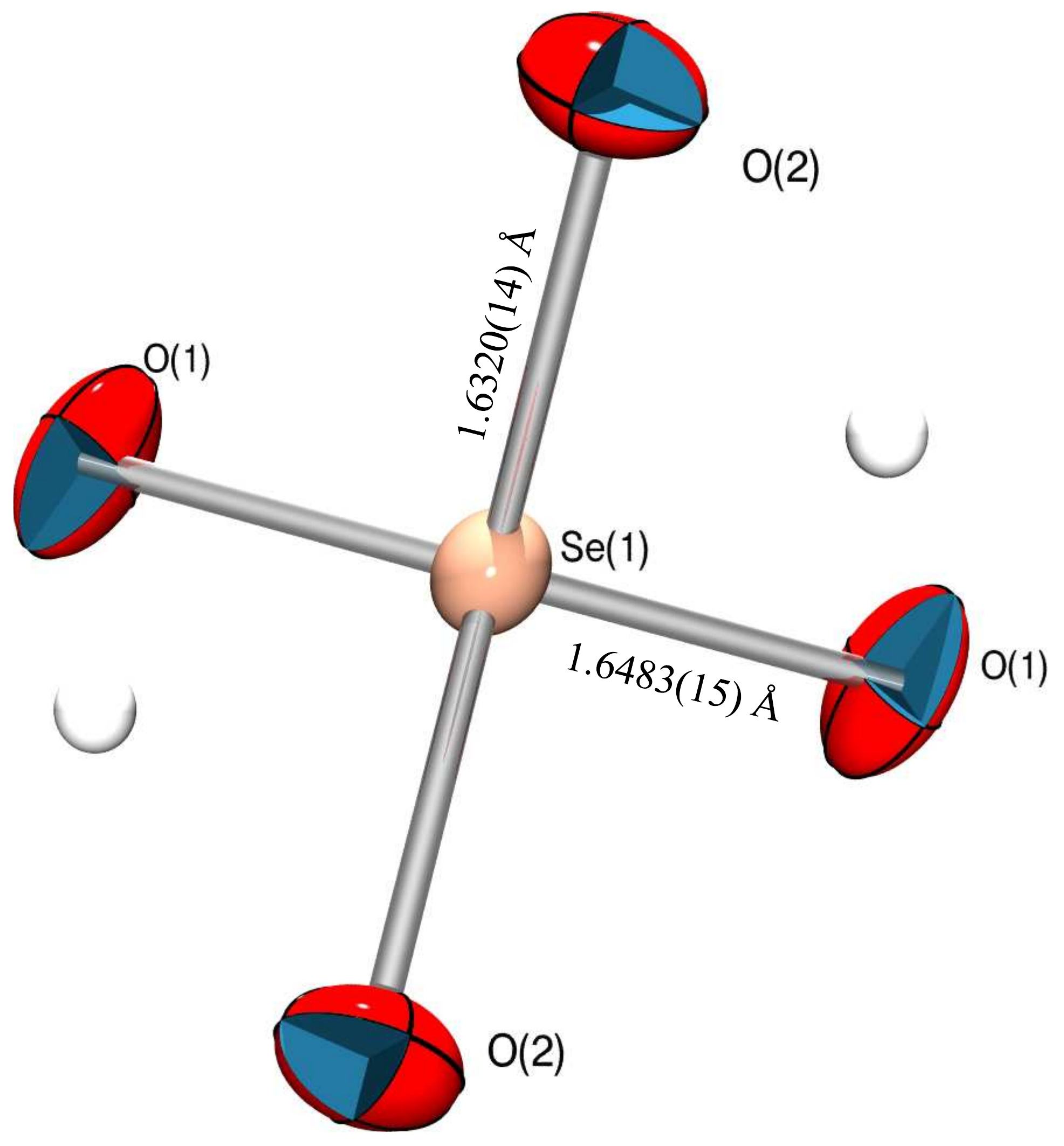}
   \caption{Selenium-oxygen interatomic bond distances for the $\mathrm{SeO_{4}^{2-}}$ anions, in $\ce{K2SeO4\cdot H2SeO3}$ crystal structure.  The figure was generated with the crystal structure visualization program Ortep III \cite{Farrugia2012} and with the aid of the program POV-Ray \cite{ThePOV-Ray2000}.}
  \label{fig:selenateanion}
 \end{center}
\end{figure}

	The selenate anions participate in two hydrogen bonds as a proton acceptor through the symmetry-equivalent O(1) and O(3) oxygens.  Due to this, the Se(l)-O(1) and  Se(l)-O(1)\texttt{\#}1 bonds (1.6484(13) \AA) are longer than the two remaining Se(l)-O(2) and Se(l)-O(2)\texttt{\#}1 (1.6319(13) \AA), as shown in Table~\ref{tbl:Se-O distances} and Figure~\ref{fig:selenateanion}.

\begin{table}[H]
\centering
  \caption{Selenium-oxygen interatomic bond distances and angles for the $\mathrm{SeO_{4}^{2-}}$ anions, in $\ce{K2SeO4\cdot H2SeO3}$ crystal structure.}
  \label{tbl:Se-O distances}
  \tabcolsep=0.11cm
  \begin{tabular}{llll}
    \hline
Bond													&Distance (\AA)	& Bond angle							 					& Angle ($\degree$)\\
	\hline
Se(1)-O(2)\textsuperscript{\#1} 						&1.6319(13)		&O(2)\textsuperscript{\#1}-Se(1)-O(2)						& 110.22(10)\\
Se(1)-O(2) 												&1.6319(13)		&O(2)\textsuperscript{\#1}-Se(1)-O(1)\textsuperscript{\#1}	& 109.23(7)	\\
Se(1)-O(1)\textsuperscript{\#1} 						&1.6484(13)		&O(2)-Se(1)-O(1)\textsuperscript{\#1}						& 108.54(7)	\\
Se(1)-O(1) 												&1.6484(13)		&O(2)\textsuperscript{\#1}-Se(1)-O(1)						& 108.54(7)	\\
\textbf{Average}										&\textbf{1.6401}&O(2)-Se(1)-O(1)											& 109.23(7)	\\
														&				&O(1)\textsuperscript{\#1}-Se(1)-O(1)						& 111.08(10)\\
														&				&\textbf{Average}						 					& \textbf{109.47}\\
O(1)-O(2)\textsuperscript{\#1}							&2.663(2)		&										 					&				\\
O(2)-O(1)												&2.674(3)		&										 					&				\\
O(1)\textsuperscript{\#1}-O(1)							&2.718(3)		&										 					&				\\
O(2)\textsuperscript{\#1}1-O(2)							&2.663(2)		&										 					&				\\
O(1)\textsuperscript{\#1}-O(1)\textsuperscript{\#1}		&2.674(3)		&										 					&				\\
	\hline
	\end{tabular}
	
	\textsuperscript{\#1} : x,-y+3/2,-z+1 
\end{table}

  The selenate oxygen atoms that are acting as hydrogen bond acceptors have longer Se(1)-O(1) bonds distances compared to those that are not participating in the hydrogen bond scheme, due to the electron density withdrawing effect of the H-bond formation.  The hydrogens decrease the charge density of the oxygen involved in the interaction, weakening the Se(1)-O(1) bonds.  The O-Se(1)-O bonds angles, listed in Table~\ref{tbl:Se-O distances}; are not significantly different, forming a little bit of a distorted tetrahedron (almost regular tetrahedron).  The mean value Se-O distance is 1.6401 \AA which is in good agreement with the average for selenate anions in some crystals member of $\ce{Me2SeO4}$ family \cite{Kalman1970, Takahashi1987}.

\begin{figure}[!htb]
 \begin{center}
  \includegraphics[scale=0.30]{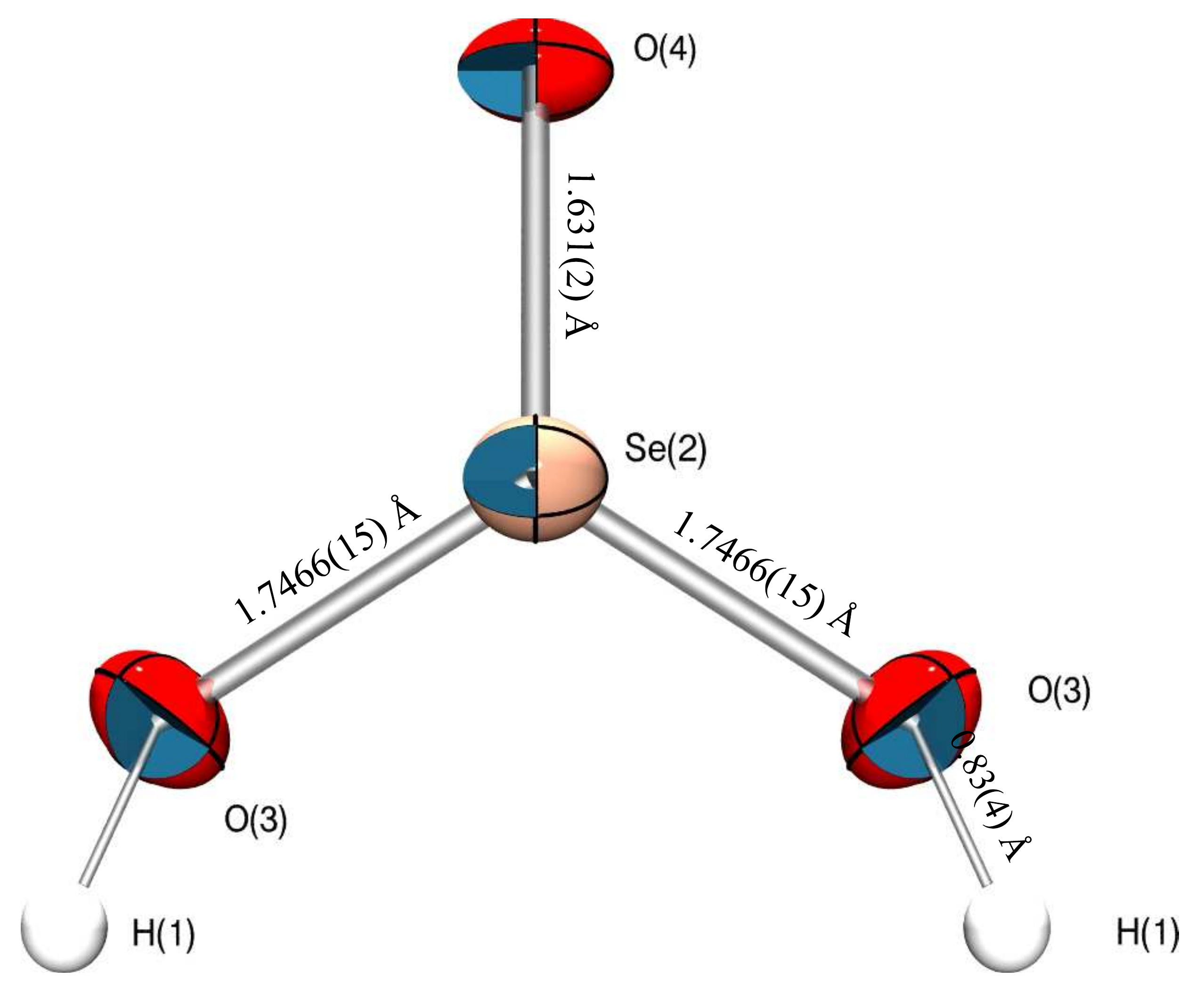}
   \caption{Selenium-oxygen interatomic bond distances and angles for the $\ce{H2SeO3}$ molecule, in $\ce{K2SeO4\cdot H2SeO3}$ crystal structure. The figure was generated with the crystal structure visualization program Ortep III \cite{Farrugia2012} and with the aid of the program POV-Ray \cite{ThePOV-Ray2000}.}
  \label{fig:H2SeO3}
 \end{center}
\end{figure}

The structure arrangement of $\ce{H2SeO3}$ molecules (see Figure~\ref{fig:H2SeO3}), is listed in Table~\ref{tbl: H2SeO3 distances}, the molecule displays an almost regular pyramidal units, joined to hydrogen atoms where the average O-Se-O angle is 98.53\degree.  

\begin{table}[H]
 \centering
  \caption{Selenium-oxygen interatomic bond distances and angles for the $\ce{H2SeO3}$ molecule, in $\ce{K2SeO4\cdot H2SeO3}$ crystal structure at 100 K.}
	\label{tbl: H2SeO3 distances}
	\tabcolsep=0.11cm
	\begin{tabular}{llll}
	\hline
	Bond						& Distance (\AA)	& Bond angle							& Angle (\degree) \\
	\hline
Se(2)-O(4) 						& 1.6311(19)		& O(4)-Se(2)-O(3)						& 98.08(7) \\
Se(2)-O(3) 						& 1.7466(15)		& O(4)-Se(2)-O(3)\textsuperscript{\#5}	& 98.08(7) \\
Se(2)-O(3)\textsuperscript{\#5} & 1.7467(15)		& O(3)-Se(2)-O(3)\textsuperscript{\#5}	& 99.43(10) \\
								&					&\textbf{Average}						& \textbf{98.53}\\

O(3)-H(1) 						& 0.83(3)			& Se(2)-O(3)-H(1)						& 115(2)	\\
			
O(4)-O(3)						& 2.552(3)			&										&			\\				
O(4)-O(3)\textsuperscript{\#5}	& 2.552(3)			&										&				\\			
O(3)-O(3)\textsuperscript{\#5}	& 2.665(3)			&										&				\\				
	\hline		
	\end{tabular}
	
	\textsuperscript{\#5}: x,y,-z+1/2
\end{table}

	The distances observed in our structure, resembles the very closely those of the crystal of selenous acid, as those reported by F. Larsen \cite{Larsen1971}.  The selenium-oxygen distances can be classified in two different categories (see the second column of Table~\ref{tbl: H2SeO3 distances}): the hydroxyl-oxygens and non-hydroxyl ones.  Here, the hydroxyl-oxygens have a longer distance to Se(2), when compared to the non-hydroxyl ones.  With Se(2)-O(4) distances of 1.631 \AA and Se(2)-O(3) distances of 1.746 \AA. The differences between Se-O(3) and Se-O(4) distances is 0.1156 \AA.  The O-H distance is 0.83(3) \AA, a short distance than that reported for $\ce{H2SeO3}$ \cite{Larsen1971}, $\ce{\beta-H2SeO3}$ \cite{Pollitt2014} and  $\ce{Na2SeO4\cdot H2SeO3\cdot H2O}$ crystals \cite{Baran1991}.

\begin{table}[H]
\fontsize{10}{12}\selectfont
 \centering
 \caption{Hydrogen bonds in $\ce{K2SeO4\cdot H2SeO3}$ crystal structure at 100 K.}
  \label{H-bonds at 100K}
  \tabcolsep=0.11cm
  \begin{tabular}{lllll}
  \hline
D-H$\cdots$A			& d(D-H) (\AA)	& d(H$\cdots$ A) (\AA)	&d(D$\cdots$A) (\AA) & Angle<(DHA)(\degree)\\
  \hline
O(3)-H(1)$\cdots$ Se(1)	& 0.83(3)		& 2.93(3)				& 3.6215(15)			& 142(3)	\\
O(3)-H(1)$\cdots$ O(1)	& 0.83(3)		& 1.78(3)				& 2.601(2)				& 171(3)	\\
\hline
 \end{tabular}
\end{table}

\begin{table}[H]
\centering
  \caption{Hydrogen coordinates (x$10^{4}$) and isotropic displacement parameters ({\AA}$^{2}$ x$10^{3}$) for $\ce{K2SeO4\cdot H2SeO3}$.  $U_{(eq)}$ is defined as one third of the trace of the orthogonalized $U_{ij}$ tensor. The figure was generated with the crystal structure visualization program Ortep III \cite{Farrugia2012} and with the aid of the program POV-Ray \cite{ThePOV-Ray2000}.}
  \label{tbl:H-coordinates}
  \tabcolsep=0.11cm
  \begin{tabular}{lllll}
    \hline
     & x 		& y 		& z 		& $U_{(eq)}$ \\
    \hline
H(1) & 5350(40)	& 6610(50)	& 3860(30)	&45(9) 		\\
\hline
  \end{tabular}
\end{table}

\begin{figure}[!htb]
\begin{center}
\includegraphics[scale=0.35]{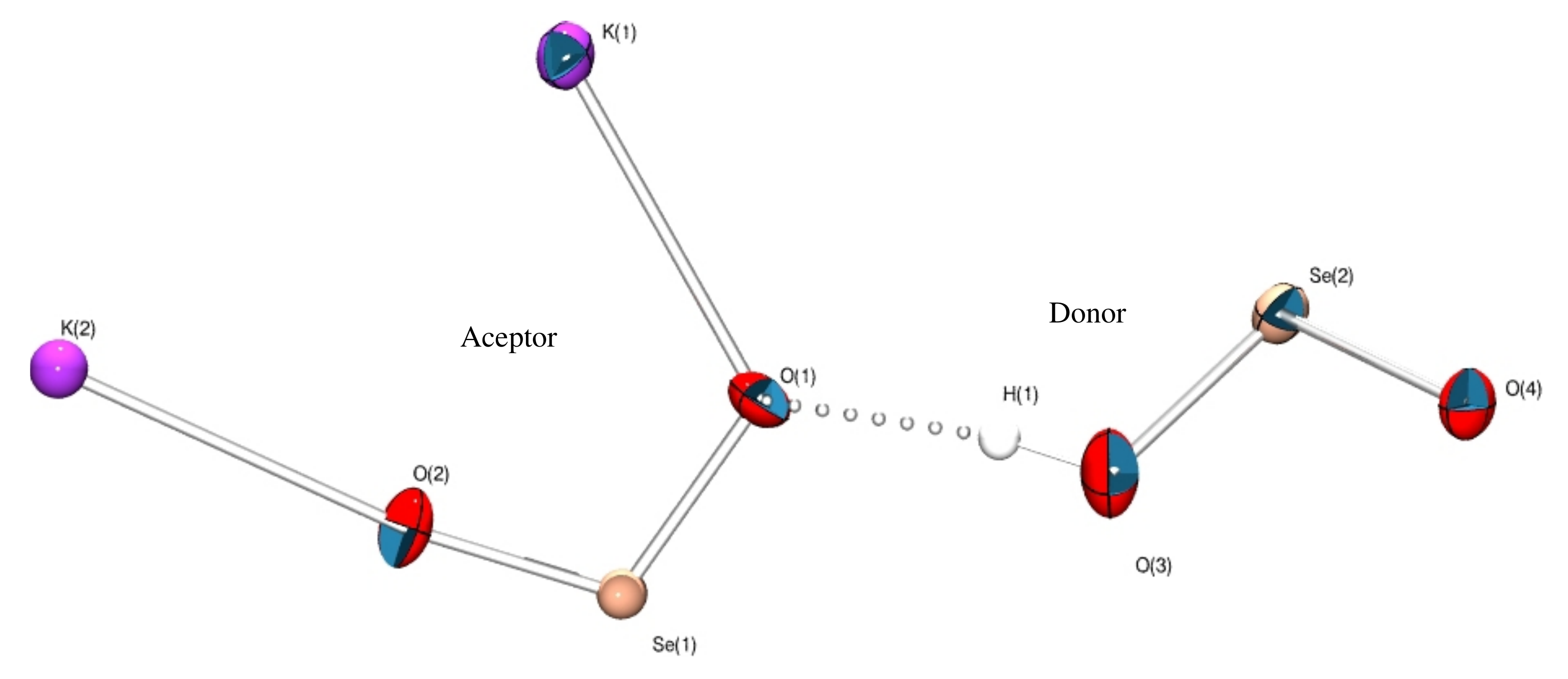}
\caption{Show the asymmetric unit of $\ce{K2SeO4\cdot H2SeO3}$.  The molecule of selenous acid $\ce{H2SeO3}$, at the right side, is labeled as Donor.   The $\mathrm{SeO_{4}^{2-}}$ anions, at the left side is labeled as Acceptor.  The figure was generated with the crystal structure visualization program Ortep III \cite{Farrugia2012} and with the aid of the program POV-Ray \cite{ThePOV-Ray2000}.}
\label{fig:Asimetric}
\end{center}
\end{figure}

The molecule of selenous acid plays the role of a proton donor in two strong hydrogen bonds, through symmetry-equivalent O(3) and O(3)\texttt{\#}5 oxygens with H(1), joined with O(1) and O(1)\texttt{\#}1 oxygens bridging with its two neighbor $\mathrm{SeO_{4}^{2-}}$ anions (see Table~\ref{H-bonds at 100K}, Table~\ref{tbl:H-coordinates} and Figure~\ref{fig:Asimetric}).

\subsection{Calculated X-ray powder diffraction information}
\begin{figure}[!htb]
\begin{center}
\includegraphics[scale=0.35]{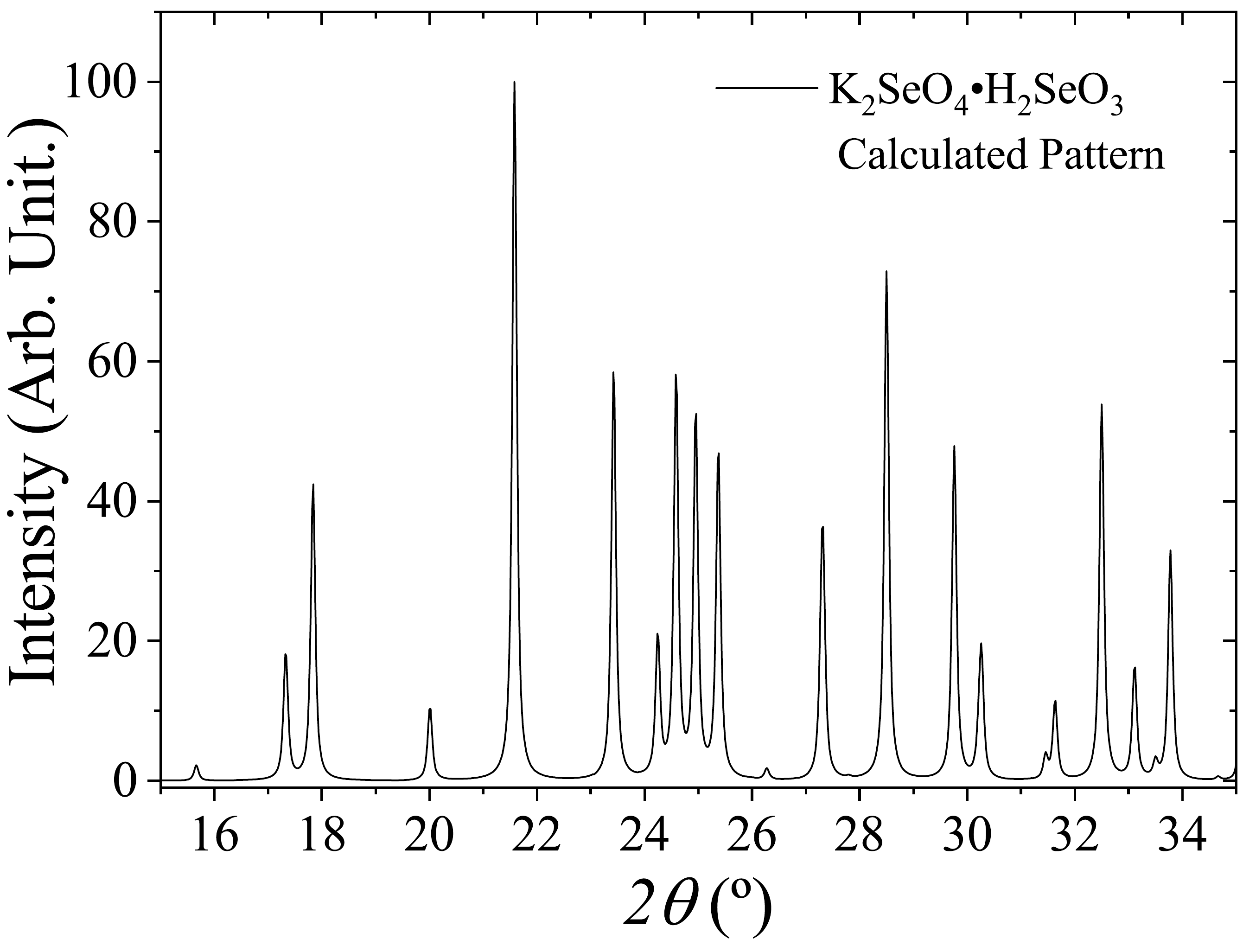}
\caption{Calculated X-ray powder diffraction pattern for $\ce{K2SeO4\cdot H2SeO3}$.}
\label{Calculated_XRPD_NS}
\end{center}
\end{figure}

	The diffractogram of the novel structure $\ce{K2SeO4\cdot H2SeO3}$, described in the previous sections, is shown in Figure~\ref{Calculated_XRPD_NS}.
	
	The Miller indices and associated parameters, related to the peaks show in Figure~\ref{Calculated_XRPD_NS} for the $\ce{K2SeO4\cdot H2SeO3}$ crystal is shown in Table~\ref{Miller indeces NS}. These results are shown  to facilatate the identification of the sample in future studies.		
	In Figure~\ref{Calculated_XRPD_NS}, a plotted of the relative intensity vs. 2$\theta$, in the range 15 to 35$\degree$, displays the calculated X-ray powder diffraction pattern, (generated with the Crystal Structure Visualisation Software Mercury [103]), of the of the compound $\ce{K2SeO4\cdot H2SeO3}$. 
		 
\begin{table}[H]
	\centering
\fontsize{10}{12}\selectfont

\caption{Miller indices and associated parameters, related to the peaks show in Figure~\ref{Calculated_XRPD_NS} for the $\ce{K2SeO4\cdot H2SeO3}$ crystal\textsuperscript{\emph{a}}. $I$: Intensity, $I_{max}$: Maximum intensity}
\label{Miller indeces NS}

\begin{tabular}{lllllllll}

\hline

2$\theta$	& $I$		& $\dfrac{I}{I_{max}}$  &$d_{hkl}$ ($d_{Bragg}$) & $\dfrac{1}{d^{2}}$	& $h$	& $k$ 	& $l$	& $\dfrac{1}{d^{2}_{Bragg}}$ \\
\hline
15.66		& 222		& 2.22					& 5.668139035			 & 0.031125692			& 1		& 1		& 0		& 0.031302297 \\
17.32		& 1814		& 18.14					& 5.128455134			 & 0.038021293			& 1		& 1		& 1		& 0.038247899 \\
17.84		& 4241		& 42.41					& 4.980128375			 & 0.040319852			& 1		& 0		& 2		& 0.040500647 \\
20			& 1024		& 10.24					& 4.446893773			 & 0.050569247			& 2		& 0		& 0		& 0.050872957 \\
21.58		& 10000		& 100					& 4.12475581			 & 0.058776472			& 1		& 1		& 2		& 0.059084705 \\
23.42		& 5848		& 58.48					& 3.804699708			 & 0.069081098			& 2		& 1		& 0		& 0.069457015 \\
24.24		& 2106		& 21.06					& 3.677822022			 & 0.073929638			& 0		& 2		& 0		& 0.074336233 \\
24.58		& 5812		& 58.12					& 3.627715498			 & 0.075985996			& 2		& 1		& 1		& 0.076402617 \\
24.96		& 5250		& 52.5					& 3.573342227			 & 0.07831605			& 2		& 0		& 2		& 0.078655365 \\
25.38		& 4688		& 46.88					& 3.515155126			 & 0.080930274			& 0		& 2		& 1		& 0.081281835 \\
26.28		& 182		& 1.82					& 3.396783842			 & 0.086669078			& 1		& 2		& 0		& 0.087054472 \\
27.32		& 3634		& 36.34					& 3.269798405			 & 0.093531531			& 1		& 1		& 3		& 0.093812715 \\
28.5		& 7292		& 72.92					& 3.137049237	 		 & 0.101614885			& 0		& 2		& 2		& 0.102118641 \\
29.76		& 4790		& 47.9					& 3.007043171			 & 0.110591226			& 0		& 0		& 4		& 0.111129632 \\
30.26		& 1968		& 19.68					& 2.958488258			 & 0.114251076			& 3		& 0		& 0		& 0.114464154 \\
31.46		& 414		& 4.14					& 2.848329986			 & 0.123259215			& 1		& 0		& 4		& 0.123847871 \\
31.64		& 1147		& 11.47					& 2.832536221			 & 0.124637594			& 2		& 2		& 0		& 0.12520919  \\
32.5		& 5386		& 53.86					& 2.759524428			 & 0.131320194			& 2		& 1		& 3		& 0.131967433 \\
33.12		& 1621		& 16.21					& 2.709273027			 & 0.136236806			& 0		& 2		& 3		& 0.136846651 \\
33.5		& 352		& 3.52					& 2.679406659			 & 0.139290897			& 3		& 1		& 1		& 0.139993814 \\
33.78		& 3298		& 32.98					& 2.657836354			 & 0.141560968			& 1		& 1		& 4		& 0.142431929 \\
34.66		& 65		& 0.65					& 2.592348003			 & 0.148803586	 		& 1	 	& 2	 	& 3		& 0.14956489  \\
\hline

\end{tabular}

\textsuperscript{\emph{a}} Crystal name: $\ce{K2SeO4\cdot H2SeO3}$, $I_{max}$: 10000, Unit Cell parameters: $a$ = 8.8672 {\AA}, $b$ = 7.3355 {\AA}, $c$ = 11.999 {\AA} $\alpha = 90\degree$, $\beta = 90\degree$, $\gamma = 90\degree$.
\end{table}

 \section{Conclusions}
	The structural arrangement of $\ce{K2SeO4\cdot H2SeO3}$ crystals resembles the selenous acid structure reported in the literature.  Instead of having the sequence ($\mathrm{H_{2}SeO_{3}-H_{2}SeO_{3})_{n}}$, the novel crytalline compound is formed by the alternating sequence molecule-anion-molecule $ \mathrm{(K_{2}SeO_{4}-H_{2}SeO_{3})_{n}}$, that is to say, the $\ce{K2SeO4}$ replaces one of the $\ce{H2SeO3}$ units.  This pattern is possible, due to the electrical charge nature of both anions.  In $\ce{K2SeO4\cdot H2SeO3}$, the charge of K$^{+}$ ions are balancing the ionic part of the structure while the hydrogen atoms are equilibrating the molecular part of the structure, see Figure~\ref{fig:comparison}.
  
	Substantial differences were not observed in the interatomic distances present in the pyramidal units of $\ce{SeO3}$ in the $\ce{H2SeO3}$ and $\ce{\beta-H2SeO3}$ crystallographic structures and the corresponding unit present in this work.  Unlike $\ce{H2SeO3}$, significant differences were found between the interatomic distances and chemical environment present in the $\ce{K2SeO4}$ structure reported and the corresponding environment in the $\ce{K2SeO4\cdot H2SeO3}$ structure. These differences could be attributed to the type of interactions (ion-ion, ion-dipole) governing the crystal packing.  
	
	The chemical environment surrounding K(1) and K(2) ions, in the structure reported herein, are different.  The potassium ion labelled K(1) has ten oxygen atoms in its neighborhood while K(2) has eight. However, both types of potassium atoms have eight oxygen atoms as their nearest neighbor, with an average potassium-oxygen distance of 2.8438 and 2.8451 {\AA}, for K(1) and K(2) respectively.  
	
	The selenium-oxygen distance present in the $\mathrm{SeO_{4}^{2-}}$ anions and those in $\ce{H2SeO3}$ can be classified into two groups, the ones where the hydrogen interacts and the ones where there are no interactions. Notice that in the selenate unit, the distances Se(1)-O(2) are slightly longer than the Se(1)-O(1), while in the selenous acid moiety, the Se(2)-O(3) distance is around 1 {\AA} larger that the Se(2)-O(4). See Table~\ref{tbl:Se-O distances}, Table~\ref{tbl: H2SeO3 distances}.

\subsection{Future work}
	It seems plausible to study the $\ce{K2SeO4\cdot H2SeO3}$ crystals at above 299 K to see if a phase transition takes places, in the compound and analyze the physical and chemical properties of the crystals at room and higher temperatures, such as non-linear optical properties and fast-proton conductivity.  One could do impedance spectroscopy measurements on the $\ce{K2SeO4\cdot H2SeO3}$ crystals to determine if it behaves as a solid state proton conductor. If it does, then it would be necessary to analyze the temperature dependence of proton conductivity along the $a$, $b$ and $c$ axis to know if the conductivity is anisotropic, since the hydrogen bonding is extended along the $c$ axis and one would expect to see higher proton conductivity.
It is very likely that $\ce{K2SeO4\cdot H2SeO3}$ structure also behaves in a similar fashion to $\ce{Na2SeO4\cdot H2SeO3\cdot H2O}$, which generates efficient second harmonic radiation at 435.8 nm (blue region of the visible spectrum).  This may require some empirical or semi- empirical calculation.

It is seeming reasonable to synthesize the $\ce{M2AO4\cdot H2AO3}$ and $\ce{M3RO4\cdot H3RO3}$ families (M = Li$^{+}$, Na$^{+}$, K$^{+}$, Rb$^{+}$, $\mathrm{(NH_{4})^{+}}$, Cs$^{+}$; A = S, Se, R= P, As), specially the $\ce{K3PO4\cdot H3PO3}$ and $\ce{Cs3PO4\cdot H3PO3}$ crystals, in order to study the new world of properties and possibilities that might arise in this tow families of compounds.
 
	Many industrial applications require the use of selenous acid ($\ce{H2SeO3}$) \cite{ScarlatoE.A.;Higa}, such as, changing the color of steel, in particular the steel in guns, but it is difficult to transport it due to its high toxicity and corrosiveness.  For this reason, it would be necessary to develop a method to separate the selenous acid from the $\ce{K2SeO4}$.  These studies suggest that perhaps, it may be safer to transport $\ce{H2SeO3}$ in the form of $\ce{K2SeO4\cdot H2SeO3}$, a polymeric crystal, instead of $\ce{H2SeO3}$ alone.
	
\section{Acknowledgement}

	Financial support of this work is acknowledged to the Center for Chemical Sensors Development at the Department of Chemistry, University of Puerto Rico-Mayagüez.

The authors thanks the Department of Chemistry, at the University of Missouri-Columbia for access to its instruments at the X-ray laboratory, for their technical assistance and expertise.
	 
In adition, thanks to the College of Arts and Sciences of UPRM, and its Dean, Prof Manuel Valdés Pizzini, for providing the traveling and financial support for the Columbia, Missouri experience.

The authors to acknowledge the Department of Chemistry, UPR-Mayagüez for providing the access to X-ray facility. 
During the course of this investigation valuable insights and very helpful discussions with retired Prof. Oscar Rosado-Lojo took place, of which I am gratefully appreciative.  We also thank the retired Prof. Rene S. Vieta for his encouragement and Sonai K. Donnell for her contribution in editing this document. 

\bibliographystyle{unsrt}

\bibliography{K2SeO4.H2SeO3}

\section{Appendix}
\begin{table}[H]
 \centering
 \caption{Symmetry transformations used to generate equivalent atoms}
  \label{Symmetry}
  \tabcolsep=0.11cm
  \begin{tabular}{lllll}
  \hline
  &    \\
  \hline
\texttt{\#}1 x,-y+3/2,-z+1		&\texttt{\#}2 -x+2,-y+1,-z+1	&\texttt{\#}3 -x+2,y+1/2,z	& \\	
\texttt{\#}4 x,y+1,z			&\texttt{\#}5 x,y,-z+1/2 		& \texttt{\#}6 -x+1,-y+1,z-1/2 	& \\
\texttt{\#}7 -x+1,-y+1,-z+1		&\texttt{\#}8 x,-y+1/2,-z+1		&\texttt{\#}9 -x+1,y-1/2,z 	& \\	
\texttt{\#}10 -x+2,y-1/2,z		&\texttt{\#}11 x,y,-z+3/2		&\texttt{\#}12 x+1,-y+3/2,-z+1	& \\
\texttt{\#}13 -x+2,y-1/2,-z+3/2	&\texttt{\#}14 -x+2,-y+1,z+1/2	&\texttt{\#}15 x-1,-y+3/2,-z+1 	& \\
\hline
\end{tabular}
\end{table}

\end{document}